\pdfoutput=1
\documentclass[a4paper,american,floatfix,pdftex,superscriptaddress,%
aps,pra,%
reprint,twocolumn,final
]{revtex4-2}
\usepackage[T1]{fontenc}
\usepackage{graphicx}
\usepackage[utf8]{inputenc}
\usepackage[ams,todos]{CMImacros}
\usepackage{xr-hyper}
\usepackage{hyperref}

\fxsetup{final}

\graphicspath{{./fig/}} 
\newcommand{\cfeldesy}{\affiliation{Center for Free-Electron Laser Science CFEL,
      Deutsches~Elektronen-Synchrotron DESY, Notkestraße~85, 22607 Hamburg, Germany}}%
\newcommand{\uhhcui}{\affiliation{Center for Ultrafast Imaging, Universität Hamburg, Luruper
      Chaussee~149, 22761~Hamburg, Germany}}%
\newcommand{\uhhphys}{\affiliation{Department of Physics, Universität Hamburg, Luruper Chaussee~149,
      22761 Hamburg, Germany}}%
\newcommand{\asemail}{\email[]{amit.samanta@cfel.de}}%
\newcommand{\cmiweb}{\homepage[URL:~]{https://www.controlled-molecule-imaging.org}}%

\begin{document}
\title{Controlled beams of cryo-cooled protein-like nanoparticles}  %
\author{Jingxuan He}\cfeldesy\uhhcui\uhhphys %
\author{Karol Długołęcki}\cfeldesy 
\author{Hubertus Bromberger}\cfeldesy 
\author{Amit K.\ Samanta}\asemail\cmiweb\cfeldesy\uhhcui
\author{Jochen Küpper}\cfeldesy\uhhcui\uhhphys%
\date{\today}%
\begin{abstract}\noindent%
   We report a cryogenic buffer-gas-cell-aerodynamic-lens-stack setup that enables the generation of
   shock-frozen, dense, and controllable beams of various nanoparticles in the gas phase, including
   small and low-density species such as isolated proteins. We demonstrate characterization of the
   setup using strong-field ionization combined with velocity-map imaging, allowing the unambiguous
   detection of nanoparticles in the protein-size range and full reconstruction of the particle
   beams including determination of particle flux and number density. The generation and
   characterization workflow presented here provides a valuable approach for protein-like sample
   preparation and delivery in single-particle diffractive imaging, microscopy, and low-temperature
   nanoscience.
\end{abstract}
\maketitle%

\section{Introduction}
Single-particle diffractive imaging (SPI) using x-ray free-electron lasers (XFELs) has emerged
as a powerful tool for obtaining structural information on various kinds of
nanoparticles~\cite{Barty:ARPC64:415, Ekeberg:PRL114:098102, Schot:natcomm6:5704,Sobolev:CommPhys3:97, Ekeberg:LSA13:15}.
In a typical SPI experiment, a dense beam of isolated and identical gas-phase nanoparticles is
continuously delivered to the focus of an x-ray beam using aerodynamic lens stacks (ALS)~\cite{Liu:AST22:293,Bogan:AST44:i,Roth:JAS124:17,Worbs:JACr54:1730,Roth:NIMA1058:168820}. In the diffract-before-destruct
approach~\cite{Neutze:Nature406:752}, diffraction patterns are recorded from every nanoparticle
before they are destroyed by the XFEL's ultra-short and extremely bright x-ray
pulse~\cite{Chapman:PTRSB369:20130313}. Millions of such diffraction patterns are required to
reconstruct the three-dimensional structure of the nanoparticle~\cite{Ayyer:Optica8:15}. The
obtained structural resolution critically relies, among other parameters, on the hitrate and
homogeneity of the nanoparticle beam~\cite{Barty:ARPC64:415, Mandl:JPCL11:6077}.

One promising and fascinating application of SPI is obtaining the high-resolution structure of
proteins without the need for crystallization~\cite{Neutze:Nature406:752, Gaffney:Science316:1444,
   Miao:Science348:530}. In particular, if proteins in their native state are irradiated by XFEL
pulses, in principle, SPI can provide native-structure information. At the current state of the art, GroEL, a 14~nm-diameter protein complex, remains the smallest biological sample successfully imaged by SPI~\cite{Ekeberg:LSA13:15}.

However, several challenges related to sample injection, among others, still remain: First, the
low inertia associated with the small size and low density of isolated proteins makes them more
susceptible to Brownian motion than nanoparticles with higher inertia~\cite{Liu:AST22:293}. This effect reduces
transmission efficiency and degrades the focusing performance of the
ALS~\cite{Liu:AST22:293, Worbs:JACr54:1730, Ekeberg:LSA13:15}, eventually resulting in low hitrate at the x-ray
focus. Additionally, proteins may deviate from their native structures in the gas phase due to dehydration caused by rapid evaporation of the surrounding water~\cite{Breuker:pnas105:18145}, thereby reducing beam homogeneity~\cite{Spoel:macrobio11:50,Meyer:WIRCMS3:408,Mandl:JPCL11:6077}.

To overcome these limitations, our group implemented a cryogenic buffer-gas cell (BGC) that can
shock-freeze nanoparticle beams by allowing them to collide with a precooled helium buffer
gas~\cite{Singh:PRA97:032704,Samanta:StructDyn7:024304}. Beams of 220~nm polystyrene nanoparticles
and granulovirus ($265\times 265 \times 445~\mathrm{nm}^{3}$) produced by the BGC were characterized
through optical-scattering localization microscopy, demonstrating high particle flux and good
controllability~\cite{Samanta:StructDyn7:024304}. In addition, calculations suggested that nanoparticles can be rapidly
cooled to 4~K upon injection into the BGC, with cooling rates ranging from $10^{4}$ to
$10^{7}$~K/s~\cite{Samanta:StructDyn7:024304}. This suppresses Brownian motion before injection into
and inside the ALS, thereby improving its transmission efficiency and focusing performance and
ultimately increasing the hitrate~\cite{Samanta:StructDyn7:024304}. Rapid cooling also helps in
retaining the native structures of proteins by cooling the sample below the glass temperature before
significant structural distortions occur and preserving the surrounding water as amorphous
ice~\cite{Al-amoudi:JSB148:131, Samanta:StructDyn7:024304}.

Previously, the performance of the BGC was only characterized using nanoparticles larger than 100~nm due
to the limitations of optical-scattering detection, as smaller particles could not be distinguished
reliably from noise without infeasibly high laser intensities~\cite{Worbs:OptExp27:36580}.
Strong-field ionization (SFI) is used as an alternative technique that allows the detection of
nanoparticles smaller than the optical-scattering limit~\cite{Powell:optexp27:27124,
   Davino:SciRep12:2045}. In a typical SFI scheme, irradiation of a nanoparticle by a femtosecond
laser pulse induces near-field enhancement at its
surface~\cite{Zherebtsov:NatPhys7:656,Powell:optexp27:27124, Seiffert:advphysx7:2010595}, generating
a substantial number of electrons that can be detected using a velocity-map imaging (VMI)
spectrometer~\cite{Eppink:RSI68:3477}. Taking advantage of this near-field enhancement, SFI can be operated at reduced laser
intensities, where nanoparticles still produce a detectable electron signal while background
ionization remains low. This makes SFI a promising approach for extending characterization to
small-nanoparticle beams.

Building on our previous BGC design~\cite{Singh:PRA97:032704,Samanta:StructDyn7:024304} and
established knowledge in ALS geometry optimization~\cite{Roth:JAS124:17, Worbs:JACr54:1730,
   Peravali:pf37:033380}, we present the design of a cryogenic
buffer-gas-cell-aerodynamic-lens-stack (BGC-ALS) setup capable of generating high-density beams of
protein-like nanoparticles, \ie, small low-density particles. Furthermore, we
characterized the performance of this setup using SFI and VMI by quantifying
key parameters such as beam width and particle flux, which are particularly beneficial for achieving
optimal hitrates in large-scale-facility experiments. This new workflow, including generating and
evaluating protein-like nanoparticle beams, provides a practical framework for future SPI
experiments with proteins.

\section{Apparatus description}
The experimental setup primarily consists of a sample injection module, the BGC-ALS, and the SFI-VMI
detection module. Both the BGC-ALS and the detection region are housed in a vacuum chamber assembled
from standard ConFlat (CF) flanges. The chamber is evacuated by a turbo-molecular pump (Pfeiffer
Vacuum HiPace 2300). An overview of the
setup is shown in \autoref[a]{fig:setup_overview}.
\begin{figure*}
   \includegraphics[width=\linewidth]{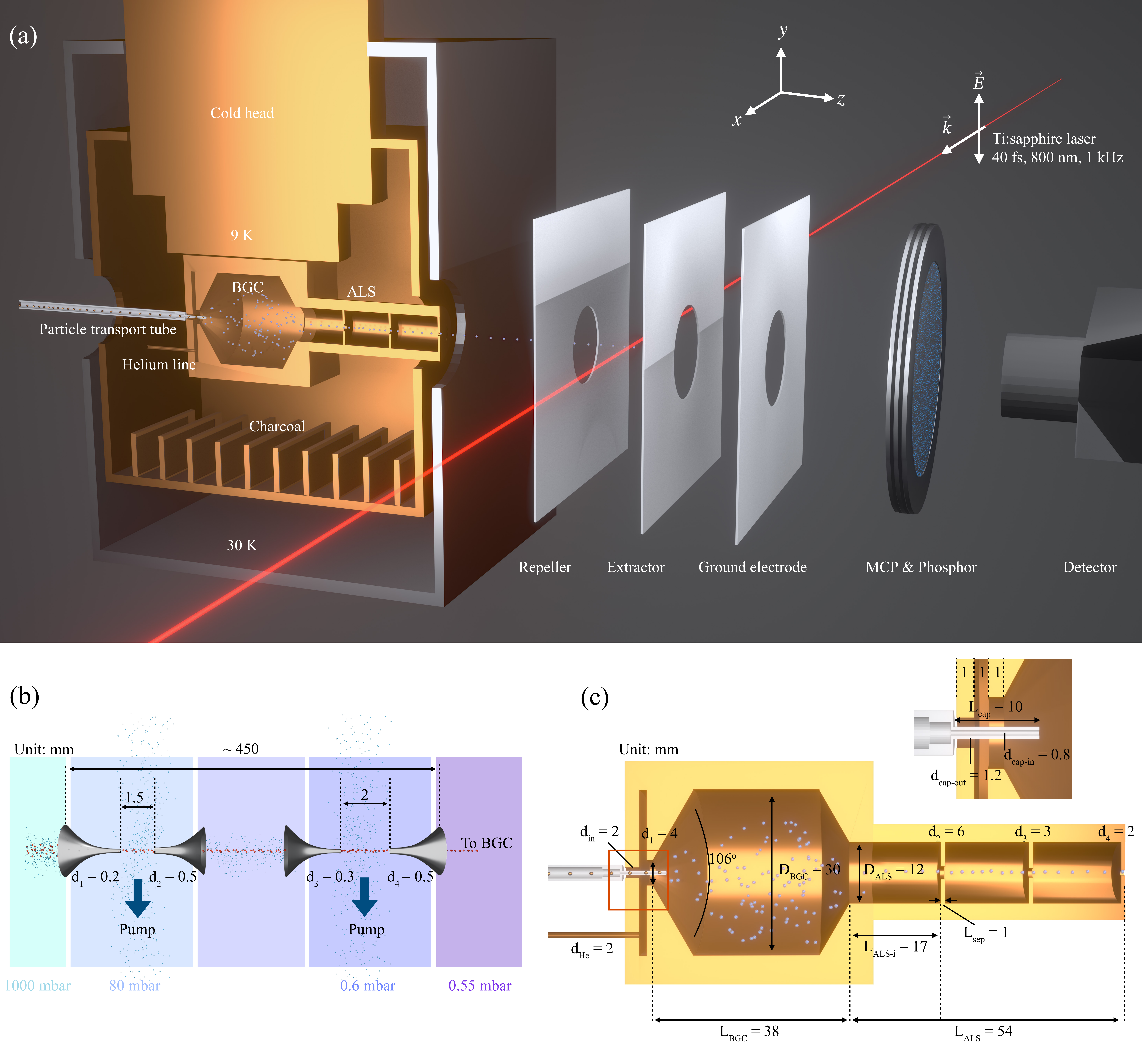}
   \caption{(a) Overview of the experimental setup. Nanoparticles are delivered through a transport
      tube into the BGC-ALS assembly, which is mounted on a two-stage cryocooler. Inside the
      BGC-ALS, nanoparticles are cooled and guided via collisions with precooled helium buffer gas.
      The resulting cold-nanoparticle beam exits the ALS and enters a VMI spectrometer in a co-axial
      geometry. A femtosecond laser beam is focused into the nanoparticle beam between the repeller
      and extractor. The ions/electrons generated by laser irradiation are projected onto a MCP/phosphor screen and recorded by a Timepix3-based detector. (b)~Schematic of the
      differentially-pumped skimmer system. The nanoparticle plume passes through this system before
      entering a transport tube that connects to the BGC-ALS. The small blue spheres represent the
      carrier gas molecules, and the large red spheres represent the nanoparticles. The orifice
      diameter of each skimmer is indicated below the corresponding skimmer. (c)~BGC-ALS dimensions: The nanoparticles are injected through the central inlet, while helium buffer gas is first introduced through the lower inlet into a small compartment before the actual BGC. Both the entrance and exit cap of the BGC have a
      \degree{106} opening angle. All ALS compartments have an inner diameter of 12~mm and are
      separated by plates of identical thickness ($\text{L}_\text{sep}=1$~mm). The orifice diameter
      of each plate is denoted above the corresponding plate in millimeters. The inset shows a
      close-up view of the BGC entrance and the capillary tip.}
   \label{fig:setup_overview}
\end{figure*}

\subsection{Nanoparticle injection}
A plume of monodisperse nanoparticles is first generated from solution by atomization or
electrospray ionization. Our workflow is compatible with both approaches. Since aerosolization
occurs at atmospheric pressure and involves carrier gases, it requires coupling to a
differentially-pumped skimmer system to reduce
background noise and prevent clogging of the BGC entrance due to ice formation. As shown in
\autoref[b]{fig:setup_overview}, the system consists of two pairs of opposing skimmers, with their
tips facing one another. The skimmers can be easily replaced with those of different orifice
diameters, and the distance between them is finely adjustable. The geometric parameters of the
skimmers used in our experiment are shown in \autoref[b]{fig:setup_overview}. The regions between
each skimmer pair are connected to a roughing pump, which continuously removes excess carrier gas.
The pressure in these regions is monitored by pressure gauges to assess the local vacuum conditions. After the last skimmer, the remaining plume primarily
contains nanoparticles and the pressure is reduced to approximately 0.55~mbar.

This sample then enters a transport tube lined with conductive silicone tubing, which suppresses the
electrostatic attraction between the nanoparticles and the tube walls, thereby improving
particle transmission. The downstream end of the transport tube is attached to a copper capillary
tip with an inner diameter of 0.8~mm, as depicted in the inset of \autoref[c]{fig:setup_overview}.
The transport tube is mounted on a three-dimensional position manipulator, enabling precise
alignment of the capillary tip with the BGC entrance. During the experiments, the capillary tip was inserted into the BGC-ALS to enhance particle transmission. Additionally, a thin heating wire operated at temperatures between 313~K and 353~K was wrapped around the capillary tip to prevent ice formation at its end in the cold environment, thereby extending the experimental runtime~\cite{Worbs:thesis:2022}.

\subsection{Cryogenic nanoparticle source}
The nanoparticle plume is then transported to the BGC-ALS. The design of the BGC section is
identical to that reported in our previous work~\cite{Singh:PRA97:032704}, except that the exit cap
is replaced with an ALS. The geometry of the ALS was optimized using CMInject
simulations~\cite{Welker:ComPhysComm:2022108138, Peravali:pf37:033380}. The resulting geometry
and dimensions of the BGC-ALS are shown in \autoref[c]{fig:setup_overview}.

As in our previous work~\cite{Singh:PRA97:032704}, the BGC-ALS is cooled by a two-stage pulsed-tube
cryocooler (Sumitomo RP082E2), which achieves temperatures of 28~K at the first stage and 3.4~K at
the second stage. As illustrated in \autoref[a]{fig:setup_overview}, an aluminum shield is thermally connected to the first stage, shielding the second stage from blackbody radiation and reducing its heat load~\cite{Singh:PRA97:032704}. The BGC-ALS is attached to the second stage and surrounded by an oxygen-free copper surface. The inner side of this copper surface is covered with charcoal flakes, which assist in the cryopumping of residual gases below 10~K~\cite{Tobin:JVSTA5:101}. Compared to the previous BGC design, the presence
of the additional ALS requires an increased shield volume, resulting in a temperature of 9~K of the
BGC-ALS.

The helium line is thermally anchored to both cooling stages using copper bobbins, allowing helium
to be sufficiently cooled before entering the BGC. The temperature mentioned in the following texts
refers to the temperature of both the BGC-ALS and the helium buffer gas, and this temperature can
be adjusted and monitored.

The precooled helium buffer gas is introduced below the nanoparticles from the same side, first into
a small compartment before the actual BGC. This design allows helium gas to rapidly fill the
compartment, enabling the formation of a dense helium flow that surrounds the nanoparticles. Inside
the BGC-ALS, the nanoparticles undergo rapid thermalization through collisions with the cold helium
atoms, thereby reducing Brownian motion. A typical helium density is $\ordsim7.9\times 10^{16}~\text{cm}^{-3}$ at a flow rate of 10~\mlnpmin~\cite{Singh:PRA97:032704}. As the nanoparticles pass through the ALS
compartments, the drag force arising from velocity difference between the nanoparticles and the
helium flow pushes them toward the central axis of the ALS, forming a cold, dense nanoparticle beam.
The beam is eventually extracted into high vacuum, with typical resulting chamber pressures of
$10^{-6}\dots10^{-5}$~mbar, while the base pressure is below $5\times10^{-9}$~mbar.

\subsection{SFI and VMI detection}
The detection and characterization of the cold nanoparticle beam are achieved by SFI combined with VMI.
The VMI spectrometer is implemented in the same vacuum chamber as the cryogenic nanoparticle source.
As illustrated in \autoref[a]{fig:setup_overview}, its design is similar to that reported by Eppink
and Parker~\cite{Eppink:RSI68:3477}, consisting of a repeller, an extractor, and a ground electrode.
Each electrode can be operated at either positive or negative voltages.

A Ti:sapphire laser (Spectra Physics, 800~nm, 40~fs, 1~kHz) is employed to ionize
nanoparticles. A linear polarizer and a half-wave plate are used to attenuate the laser intensity to
a desired value. The laser beam propagates along the $x$-direction, midway between the repeller and
the extractor, and is polarized along the $y$-direction. A 500~mm lens focuses the laser beam to a
spot with a diameter of about 50~\um, resulting in a peak intensity of approximately
$7.7\times10^{12}$~\Wpcmcm, as determined from the xenon above-threshold ionization spectrum, see
Figure S1 in the Supplementary Material. The focusing lens is mounted on a
translational stage enabling the laser focus to be adjusted along the $x$- and $y$-axis. Due to
mechanical constraints, the laser focus is positioned approximately 30~mm downstream of the ALS exit
along the $z$-axis. This distance is due to the 20~mm separation between the laser focus and the end of the ceramic insulation rod mounted on the repeller, as well as an additional 10~mm of clearance required during installation to avoid mechanical contact.

Ions and electrons generated from nanoparticles by SFI are projected by the VMI electric field onto
a microchannel plate detector (MCP, Beam Imaging Solutions BOS-40) equipped with a P-47 phosphor
screen. The active area of the MCP has a diameter of 40~mm. In the present geometry, the distance
from the laser focus to the MCP surface is approximately 155~mm. The light pulses emitted from the phosphor screen are recorded by a detector system positioned outside the vacuum chamber. This detector system, which primarily consists of an optical imaging system and a Timepix3-based sensor~\cite{Bromberger:JINST19:P11008} and is controlled by the open-source software PymePix~\cite{AlRefaie:JINST14:P10003, pymepix:gitlab}, enables the simultaneous recording of position and time-of-flight (TOF) information~\cite{Bromberger:JINST19:P11008}.

\section{Characterization of the setup}
Spherical polystyrene nanoparticles (Thermo Scientific 3020A, $d=20$~nm) were used as test samples
to characterize the performance of the setup. Their density is similar to that of proteins. To the
best of our knowledge, 20~nm is the lower practical size limit for commercially available
monodisperse polystyrene nanoparticles. The nanoparticle suspension was prepared at a concentration
of $2.84\times10^{13}$~particles/mL in ultrapure water. In this work, an atomizer (TSI 3076) was used for aerosolization. $\text{N}_{2}$ was used as the carrier gas at a backing pressure of 1.5~bar, and the sample flow rate was set to 0.2~mL/min and controlled by a syringe pump (KD Scientific 100).

\begin{figure}
   \includegraphics[width=\linewidth]{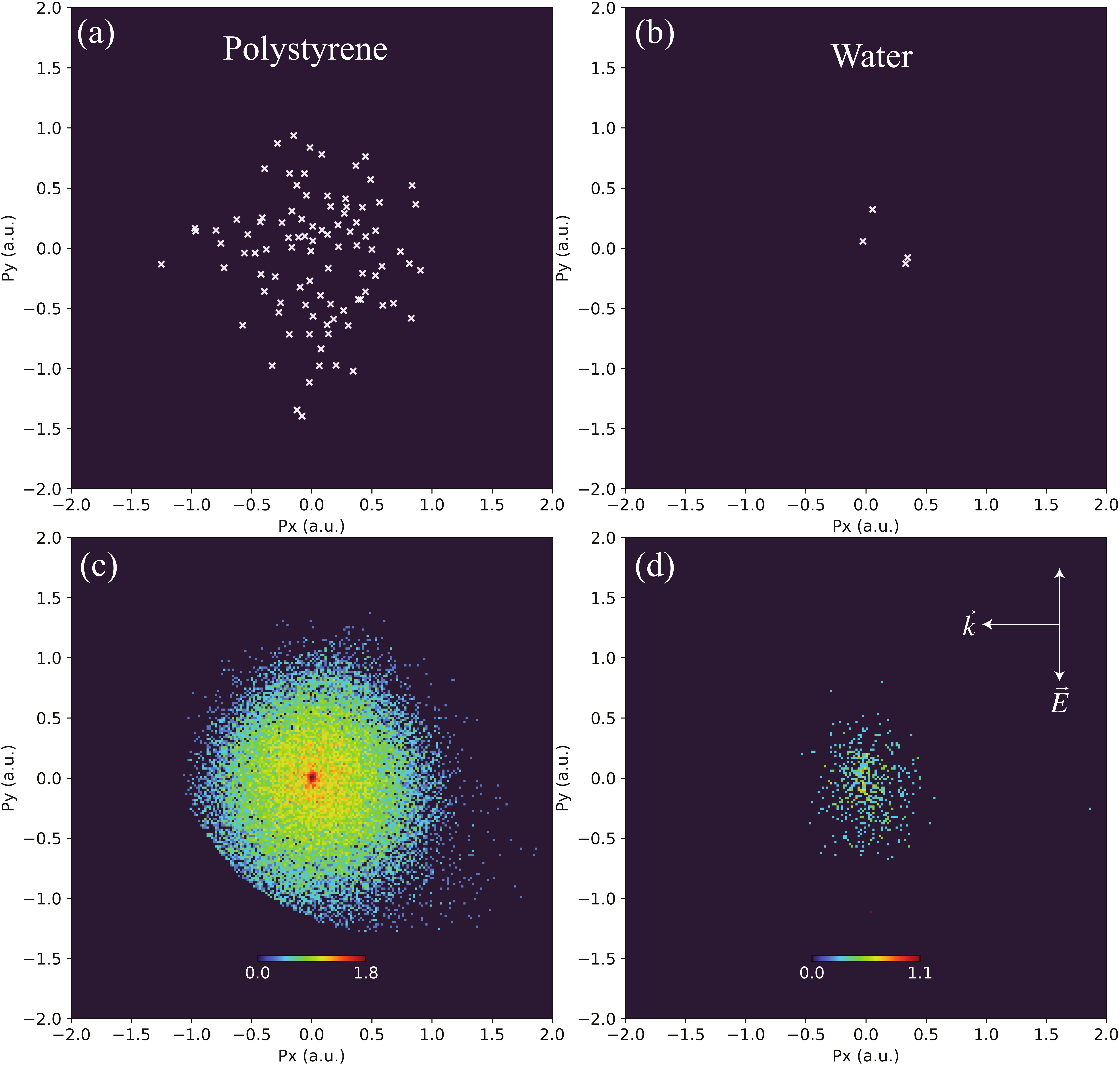}
   \caption{Typical single-shot electron VMI images obtained from (a) aerosols containing 20~nm
      polystyrene nanoparticles and (b) water vapor. Electron hits are denoted by white crosses.
      (c,d)~Corresponding two-dimensional histograms recorded over $8.4\times 10^{5}$ laser shots
      for polystyrene nanoparticles and $9.0\times10^{5}$ laser shots for water, respectively. The
      color bars indicate the electron count on a logarithmic scale. For visual clarity, the upper limit of the color scale in (d) was reduced. All images are shown in the
      atomic unit of momentum. }
   \label{fig:2dhist_electron}
\end{figure}

\begin{figure}
   \includegraphics[width=\linewidth]{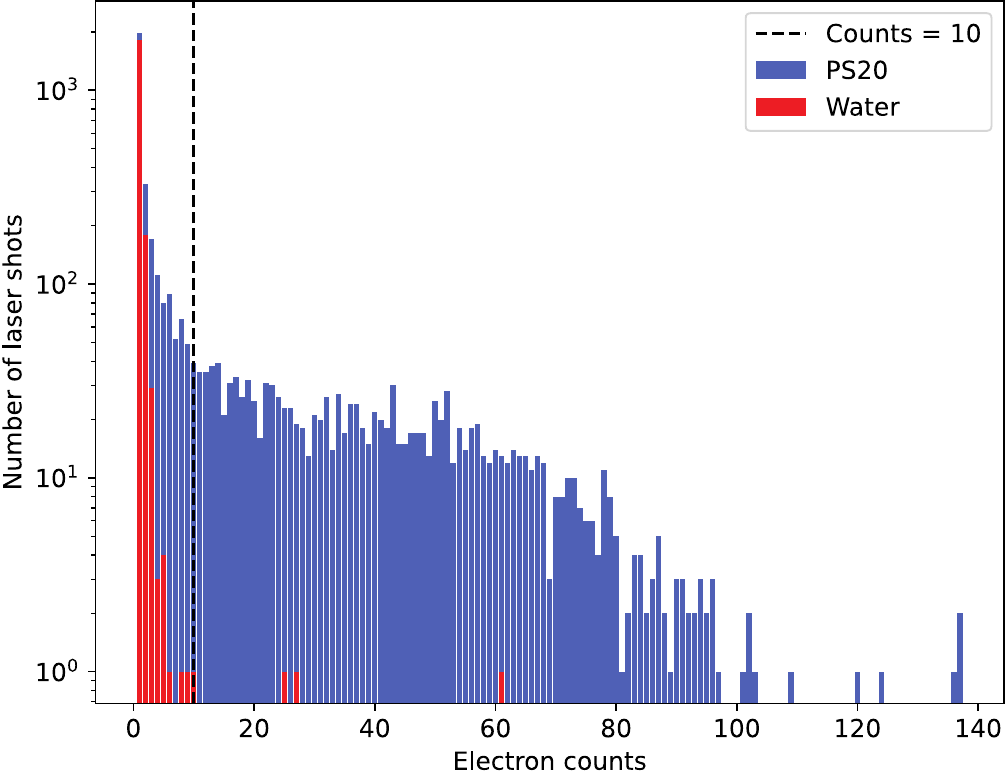}
   \caption{One-dimensional histogram of the number of detected electrons per laser shot for SFI of
      20~nm polystyrene nanoparticles (blue) and water aerosols (red), constructed from the same
      data as shown in \autoref[c,d]{fig:2dhist_electron}. The vertical dashed line indicates the
      threshold used to identify nanoparticle events. }
   \label{fig:electron_signal_compare}
\end{figure}

The VMI spectrometer was operated at negative voltages to enable detection of electrons.
Representative VMI images obtained from the nanoparticles and the water vapor generated under
identical conditions are shown in \autoref{fig:2dhist_electron}.
The recorded images were centroided using PymePix~\cite{AlRefaie:JINST14:P10003, Bromberger:JPB55:144001, pymepix:gitlab} to enable the identification
of individual electron hits and reliable hit counting. \autoref[a,b]{fig:2dhist_electron} show
typical single-shot VMI images, reflecting the two-dimensional spatial distribution of electrons
generated by a single laser shot. \autoref[c,d]{fig:2dhist_electron} present the resulting
histograms of electron hits accumulated over $8.4\times10^{5}$ (nanoparticle) and $9.0\times10^{5}$
(water) laser shots. In both cases, electrons exhibit a broad angular distribution, with slightly
higher yields along the laser polarization axis. Notably, electrons generated from nanoparticles
show a broader and higher momentum distribution than those from water vapor, with significantly
higher electron counts. To quantify this difference, the distributions of electron counts per laser
shot for nanoparticles and pure water were analyzed, as shown in
\autoref{fig:electron_signal_compare}.
To reduce interference from background noise, only laser shots yielding at least two detected electrons were included in the analysis.

For nanoparticles, a significant fraction of laser shots produce more than ten electrons, whereas
practically all laser shots produce fewer than ten electrons when only water vapor was injected. This striking contrast is consistent with previous
results~\cite{Davino:SciRep12:2045} and the three exceptions observed under water-only conditions
are most likely caused by residual nanoparticles. Given their extremely low occurrence (3 out of
$9.0\times10^{5}$), these events are statistically negligible. Consequently, laser shots yielding
ten or more electrons can be unambiguously assigned to nanoparticle events.

Assuming that each laser shot interacts with at most one nanoparticle, the particle hitrate is
therefore defined as
\begin{equation*}
   \text{hitrate} = \frac{N_{\geq10\;e^{-}}}{N_{\text{tot}}}.
\end{equation*}
Here, $N_{\geq 10\,e^{-}}$ denotes the number of laser shots producing at least ten electrons, and
$N_{\text{tot}}$ the total number of laser shots. This value represents a lower-bound estimate of
the particle hitrate, as nanoparticles may interact with the edge of the laser focus and thus
produce fewer electrons.

A small fraction of the electrons shown in \autoref[c]{fig:2dhist_electron} falls outside the active
area of the MCP, due to the slight misalignment between the central axes of the BGC-ALS and the VMI
spectrometer. Nevertheless, the majority of the electrons are recorded and the contrast between
nanoparticle and background events remains high. Therefore, the particle beam characterization is
unaffected. In the future, this issue could be resolved by using an MCP with a larger active area.

\begin{figure}
   \includegraphics[width=\linewidth]{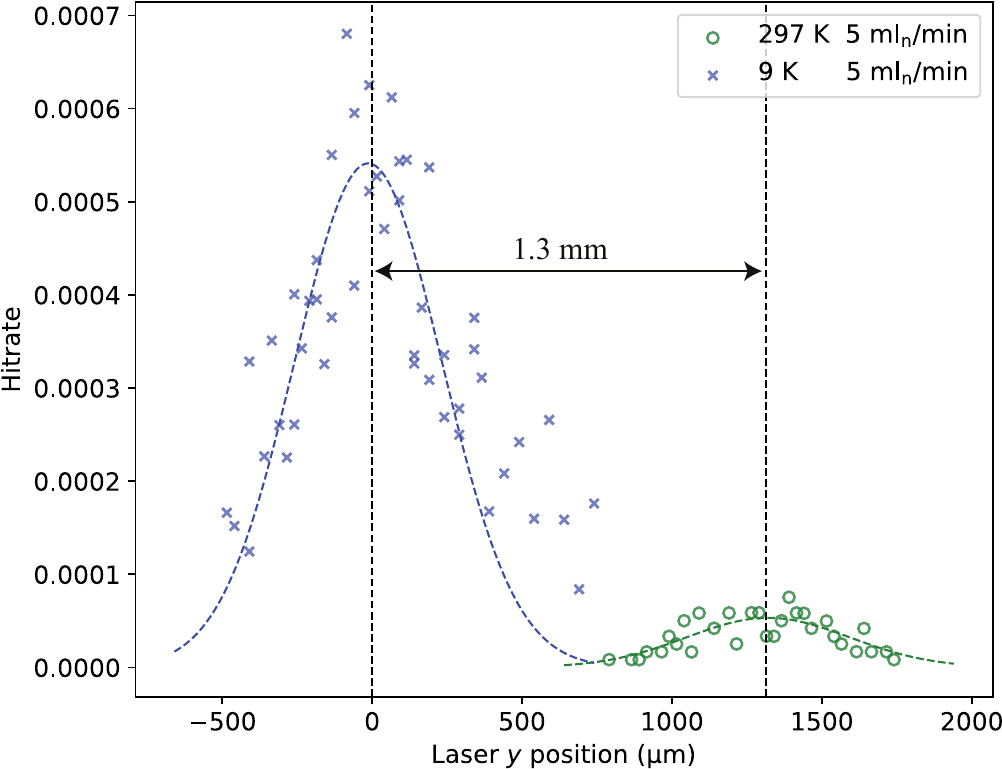}%
   \caption{Transverse profiles of 20~nm polystyrene beams recorded at room temperature (green
      circles) and 9~K (blue crosses), with helium flow rate operated at 5~\mlnpmin for both cases.
      The horizontal axis denotes the relative laser focus position along the $y$-direction, with
      smaller values corresponding to higher positions in the laboratory frame. Hitrates were
      obtained from 120~s measurements ($1.2\times10^{5}$ laser shots) at each laser position.
      Gaussian fits were applied for visual clarity and quantifying beam parameters. Two black dashed
      lines denote the center of nanoparticle beams.}
   \label{fig:temperature_shift}
\end{figure}

The transverse profile of the nanoparticle beam was measured by scanning the laser focus along the
$y$-axis and recording the hitrate at different positions~\cite{Chang:IRPC34:557,Johny:CPL721:149}.
\autoref{fig:temperature_shift} compares the transverse profiles recorded at room temperature and at
9~K with a helium flow rate of 5~\mlnpmin for both cases. When the BGC-ALS was cooled from room
temperature to 9~K, the transport tube had to be raised by approximately 1.2~mm to realign with the
BGC entrance, indicating the mechanical displacement of the BGC-ALS system due to thermal
contraction of the cold system. Consistently, the center of the measured particle beam also shifted
upward by approximately 1.3~mm. This agreement confirms that the recorded transverse profile
accurately represents the actual nanoparticle beams produced by the BGC-ALS.

Furthermore, given the high cooling rate achieved by the BGC alone~\cite{Samanta:StructDyn7:024304},
an extended thermalization region provided by the BGC-ALS allows us to safely assume that the
nanoparticles are fully thermalized to 9~K. Additionally, throughout the full range of the
transverse profile, the hitrate at 9~K is about an order of magnitude higher than that at room
temperature. This increase can be reasonably attributed to the suppression of Brownian motion, which
further confirms the cooling of the nanoparticles. When Brownian-motion-driven diffusion is reduced,
nanoparticles are less likely to reach the cell walls and be lost. As a result, they can be more
efficiently extracted by the helium flow toward the BGC-ALS exit. A more detailed analysis of
Brownian-motion reduction will be addressed in a future publication~\cite{He:BrownianMotion:inprep}.
The full width at half maximum (FWHM) was obtained from a Gaussian fit to the measured profile, and used to estimate the particle flux. The beam obtained at 9~K with helium flow rate of 5~\mlnpmin has a FWHM
of 578~\um and an estimated flux of $4.4\times10^{5}$~$\um^{-2}\,\text{s}^{-1}$. Both the FWHM and the particle flux can be
further optimized by adjusting the helium flow rate~\cite{Samanta:StructDyn7:024304,
   Worbs:thesis:2022, Peravali:cryo-als:inprep}. It should be noted that the reported FWHM and
particle flux here correspond to the measurements taken at 30~mm downstream from the ALS exit,
rather than at the actual particle beam focus, which is typically a few millimeters from the ALS
exit. Using numerical simulations, it is possible to estimate the position and FWHM of the actual
spatial focus of the particle beam~\cite{Peravali:pf37:033380, Peravali:thesis:2025}. A detailed
discussion is beyond the scope of the present work. Nevertheless, these measurements provide a
valuable reference for evaluating and optimizing particle beams.

\begin{figure}
   \includegraphics[width=\linewidth]{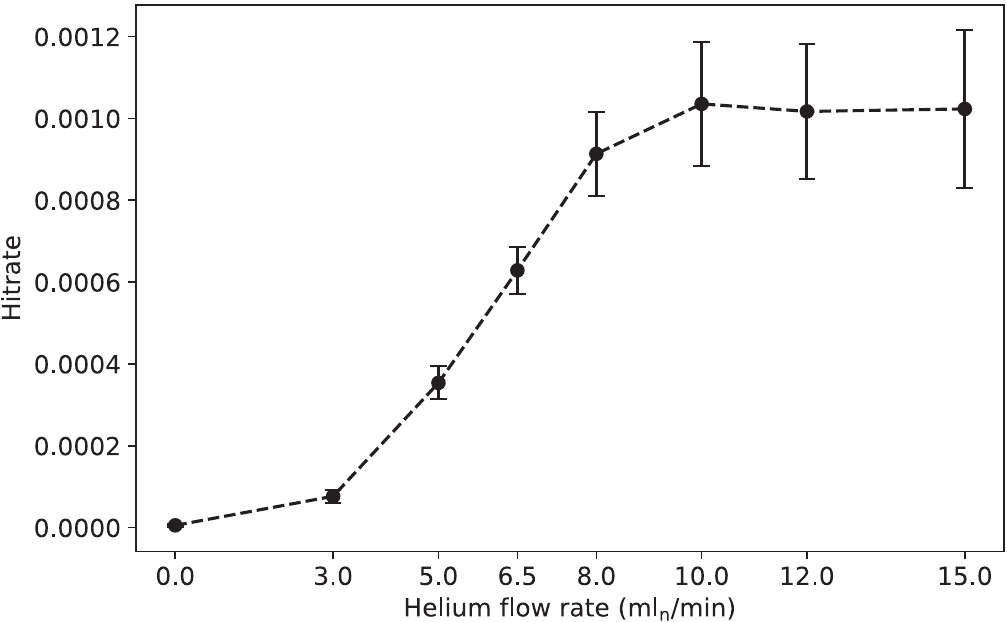}
   \caption{Averaged hitrate of 20~nm polystyrene nanoparticles as a function of helium flow rate.
      Error bars represent the standard error of the mean. }
   \label{fig:helium_scan}
\end{figure}

The influence of the helium flow rate on nanoparticle extraction is shown in
\autoref{fig:helium_scan}.
All measurements were performed at the laser vertical position corresponding to the highest particle
yield. For each helium flow rate, data were recorded for 120~s, corresponding to $1.2\times10^{5}$
laser shots. To avoid bias, the helium flow rate was first scanned in ascending order and then in
descending order, and this scan sequence was repeated eight times.
In total, 16 measurements were obtained for each flow rate. The hitrates presented in the figure
were obtained by averaging all measurements.

The hitrate is nearly zero when no helium is flowing inside the BGC-ALS. As the helium flow rate
increases from 3 to 8~\mlnpmin, the hitrate rises approximately linearly. The highest hitrate is
observed at a flow rate around 10\ldots12~\mlnpmin, followed by a plateau. The hitrates measured at
flow rates above 10~\mlnpmin exhibit relatively large error bars, primarily due to the occasional
clogging at the BGC entrance as a result of ice formation. Overall, the performance of the 20~nm
polystyrene beam is consistent with previous observations for 220~nm nanoparticles produced by the
BGC~\cite{Samanta:StructDyn7:024304}, except that the 20~nm polystyrene requires much lower helium
flow rates for extraction.

\begin{figure*}
   \includegraphics[width=\linewidth]{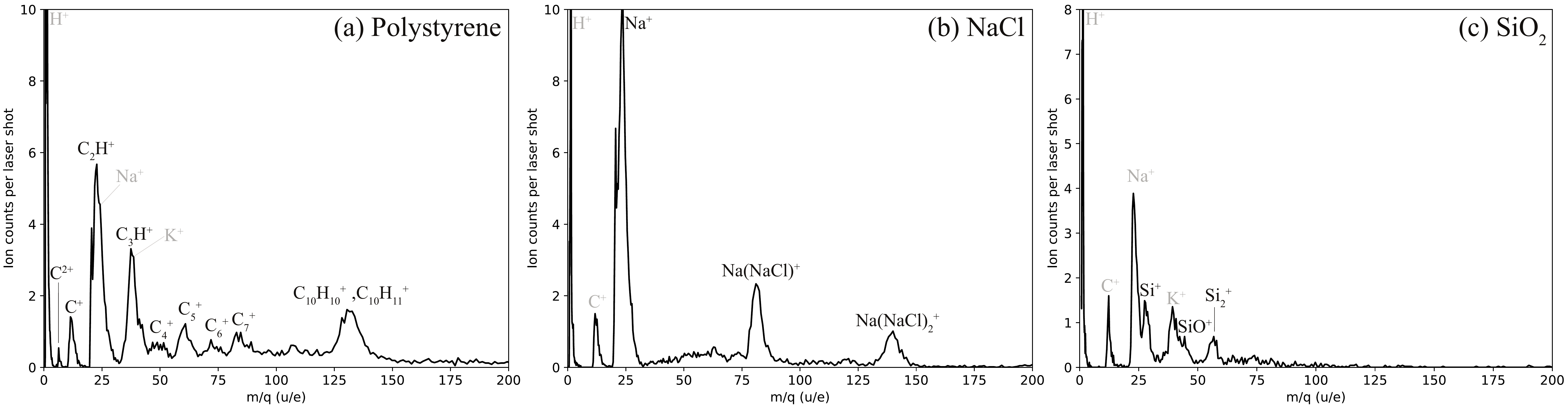}
   \caption{TOF-MS of various nanoparticles: (a)~Polystyrene, (b)~NaCl, (c)~$\text{SiO}_{2}$. In all
      cases, the helium flow was fixed at 8~\mlnpmin, the temperature at 9~K, and the laser $y$
      position at 0~\um. More details regarding samples are provided in \autoref{tab:samples}. Peaks
      assigned to background ions, \eg, carrier gas and trace-gas species, are labeled in gray.}
   \label{fig:tof-ms}
\end{figure*}

The results above demonstrate that the BGC-ALS is capable of generating shock-frozen, dense, and
controllable beams of 20~nm polystyrene nanoparticles. Compared to earlier ALS studies primarily
focused on delivering high-inertia nanoparticles, the present work extends efficient ALS-based
sample delivery to protein-like nanoparticles and demonstrates the detection of such small and light
nanoparticles that were previously nearly impossible to observe using conventional optical scattering techniques.
Although 20~nm polystyrene nanoparticles serve as a suitable model for protein-like nanoparticles,
it is also important to evaluate the general applicability of this workflow to other nanoparticle
species. Therefore, NaCl and $\text{SiO}_{2}$ nanoparticles were injected into the BGC-ALS, and the
corresponding sample properties are listed in \autoref{tab:samples}.
\begin{table}
   \centering
   \resizebox{\linewidth}{!}{
      \begin{tabular}{lccc}
        \hline
        Sample & Diameter (nm) & Laser intensity (\Wpcmcm) & Concentration (g/L) \\
        \hline
        Polystyrene & 20 & $7.7\times 10^{12}$ & 0.13 \\
        NaCl & $\approx 40$ & $1.3\times 10^{13}$ & 1 \\
        $\text{SiO}_{2}$ & 50 & $1.3\times 10^{13}$ & 0.2 \\
        \hline
      \end{tabular}
   }
   \caption{Summary of the sample properties and experimental parameters for the nanoparticles tested by TOF-MS.}
   \label{tab:samples}
\end{table}
For these tests, the polarity of the VMI electrodes was switched to positive to allow the recording
of time-of-flight mass spectra (TOF-MS), enabling the identification of particle species. As shown
in \autoref{fig:tof-ms},
distinct TOF-MS spectra were observed for different nanoparticle species. Apart from a few shared
peaks that may originate from background species, each spectrum presents characteristic
mass-to-charge ($m/q$) peaks that allow straightforward species identification. For instance, the
spectrum of polystyrene nanoparticles is dominated by $\text{C}_{m}\text{H}_{n}$ ($m\geq1, n\geq0$)
fragments, while NaCl nanoparticles exhibit characteristic NaCl cluster peaks, and $\text{SiO}_{2}$
nanoparticles yield silicon-related fragment ions. This confirms that the detected signals indeed
originate from those nanoparticles injected into the BGC-ALS. Furthermore, it demonstrates the broad
applicability of our workflow to different nanoparticle species.

The vacuum chamber containing the BGC-ALS is entirely constructed from standard CF flanges, making
it compatible with existing source chambers at large-scale facilities such as the European XFEL~\cite{Mancuso:JSR26:660,Round:JSR33:198}. As
a result, nanoparticle beams optimized under laboratory conditions can be readily reproduced at
large facilities, substantially improving the feasibility of SPI experiments. In particular, the
absence of VMI-related mechanical restrictions in SPI experiments enables the x-ray beam to be
aligned with the actual spatial focus of the particle beam, resulting in a significantly higher
hitrate. At the same time, the high-density helium introduced by BGC-ALS allows the dilution of the
carrier gases from aerosol generators ($\text{N}_{2}$, $\text{CO}_{2}$). Such dilution was shown to
reduce background elastic scattering in SPI~\cite{Yenupuri:scirep14:4401}, thus facilitating the
imaging of weak scatterers such as proteins~\cite{Ekeberg:LSA13:15, Yenupuri:scirep14:4401}.

\section{Conclusion}
We demonstrated the generation of shock-frozen, dense, and highly controllable beams of protein-like
nanoparticles using the BGC-ALS, together with reliable beam characterization through SFI and VMI.
Crucially, the high cooling rates achieved in the BGC-ALS are promising for retaining the native-like structures of biological particles. In addition, the BGC-ALS provides a versatile platform that can be applied to particles with minimal constraints on size and density.

The presented workflow enables laboratory-based optimization of protein-like nanoparticle beams and
provides a practical pathway for sample preparation prior to SPI experiments. Furthermore, in the
future, modifications of the BGC-ALS designed by numerical simulations can be readily validated
experimentally using the established workflow, enabling iterative improvement toward more focused
and denser beams of targeted particle species. Taken together, these capabilities pave the way for
supporting future three-dimensional structure determination of protein-like nanoparticles by SPI.

Moreover, the BGC-ALS can be readily adapted to a wide range of future applications. For example, it
could be employed for soft landing in cryogenic electron microscopy~\cite{Esser:sciadv10:eadl4628},
studies of neutral aerosols in the atmospheric science~\cite{Schmale:natcc11:95}, as well as
controlled coating processes in materials science~\cite{Lai:scirep13:4709}. These potential
applications of the BGC-ALS make it a promising tool for connecting fundamental and applied science
across research fields.

\section{Acknowledgments}
This work was supported by Deutsches Elektronen-Synchrotron (DESY), a member of the Helmholtz
Association (HGF) and the Cluster of Excellence ``Advanced Imaging of
Matter'' of the Deutsche Forschungsgemeinschaft (DFG, AIM, EXC~2056, ID~390715994).

\section{Data availability}
The data that support the findings of this study are available from the corresponding author upon
reasonable request.

\bibliography{string,cmi}%

\onecolumngrid%
\listofnotes%
\end{document}